\documentstyle[eqsecnum,aps]{revtex}
\input{epsf}
\begin{document}
\draft
\title{Wavelength limits on isobaricity of perturbations\\
in a thermally unstable radiatively cooling medium
}
\author{I. G. Kovalenko
}
\address{
Department of Physics, Volgograd State University, 
Volgograd 400062, Russia
}
\author{Yu. A. Shchekinov}
\address{
Department of Physics, 
Rostov State University, Rostov on Don 344090, Russia
}
\date{\today}
\maketitle


\let \pat=\partial
\let \d=\delta
\let \td=\tilde
\def \dt{{\pat\over\pat t}}
\def \dm{{\pat\over\pat m}}
\def \ddx{{d\over dx}}
\def \ddr{{d\over dr}}

\def\and{{\rm ~and}}
\def\12{{1 \over 2}}
\def\etal{{\it et~al.\ }}
\def\eg{{\it e.g.,~}}
\def\ie{{\it i.e.,~}}
\def\de{\partial}
\def\gtsima{$\; \buildrel > \over \sim \;$}
\def\simgt{\lower.5ex\hbox{\gtsima}}
\def\ltsima{$\; \buildrel < \over \sim \;$}
\def\simlt{\lower.5ex\hbox{\ltsima}}
\def\ref{\noindent\hangindent.5in\hangafter=1}

\begin{abstract}
Nonlinear evolution of one-dimensional planar perturbations 
in an optically thin radiatively cooling medium in the long-wavelength limit 
is studied numerically. The accepted cooling function generates in thermal 
equilibrium a bistable equation of state $P(\rho)$. 
The unperturbed state is taken close to the upper (low-density) 
unstable state with infinite compressibility ($dP/d\rho= 0$).  
The evolution is shown to proceed in three different stages. At first stage, 
pressure and density set in the equilibrium equation of state, 
and velocity profile steepens gradually as in case
of pressure-free flows. At second stage, those regions of the flow 
where anomalous pressure (\ie with negative compressibility) holds, 
create velocity profile more sharp
than in pressure-free case, which in turn results in formation of a very 
narrow (short-wavelength) region where gas separates the equilibrium equation 
of state and pressure equilibrium sets in rapidly. 
On this stage, variation in pressure between narrow dense region and
extended environment does not exceed more than 0.01 of the unperturbed
value. On third stage, gas in the short-wavelength region reaches 
the second (high-density) stable state, and pressure balance establishes 
through the flow with pressure equal to the one in the unperturbed state. 
In external (long-wavelength) regions, gas forms slow isobaric inflow toward 
the short-wavelength layer. The duration of these stages 
decreases when the ratio of the acoustic time to the radiative cooling time 
increases.  Limits in which nonlinear evolution of thermally unstable
long-wavelength perturbations develops in isobaric regime are obtained. 
\end{abstract}

\vfill\eject
\section{INTRODUCTION}

Thermal instability is thought to be a mechanism which 
causes a radiatively cooling medium (RCM) to break onto two phases: relatively 
dense cold filaments and clumps surrounded by rare hot plasma. 
This phenomenon is common for both cosmic and laboratory optically 
thin plasmas which lose their energy radiatively (see for more discussion
\cite{meer96}). 
A comprehensive linear theory of thermal instability has been developed in the 
pioneering paper by Field\cite{field65}. The instability results from the fact 
that the 
radiative energy loss rate grows with density. A local increase of the density 
brought in a uniform equilibrium plasma, enhances cooling rate and stimulate 
temperature to fall. This leads in turn to subsequent progressive growth of 
the density in order to keep pressure constant. 

Last decade, nonlinear aspects 
of thermal instability have been extensively investigated both analytically and 
numerically, and considerable progress in understanding of the dynamics 
of phase transitions and structure formation in bistable media was achieved. 

The equations governing dynamics of RCM, besides the 
nonlinearities intrinsically connected with gas dynamics processes, contain 
also nonlinearities related to radiative energy loss rate, 
$\dot E(\rho,T)$, which is square function of density and, as a rule, 
strongly nonlinear function of temperature. In general, nonlinear 
dynamics of such media is complex for analytical study, and 
several approximations were used to reduce the equations to more simple and 
useful form. This procedure is based on the presence of different 
characteristic time scales in RCM: characteristic radiative cooling 
time $\tau_R$, conductive time $\tau_\chi$, and acoustic time $\tau_A$. When 
these times differ considerably from each other one can find the fastest 
process which relaxes to its equilibrium state, and thus eliminate one of the 
governing equations. In this manner Doroshkevich and Zeldovich\cite{dor81}, 
assuming 
the medium to follow isobaric dynamical regime, have described dynamics of RCM  
by a nonlinear heat-diffusion equation 

$${\de T\over \de t}={\de \over \de q}\chi(T){\de T\over \de q}-
\varepsilon(T),$$
which belongs to a class of diffusion-reaction equations, here the reduced 
cooling function $\varepsilon(T)$ represents the ``reaction'' term. 
In subsequent papers Meerson and 
Sasorov\cite{meer87}, and then Meerson \etal\cite{meer93} have developed this 
approach to 
reduce the dynamics of RCM to more simple set of equations depending on 
interrelation between characteristic times. In the short-wavelength limit when 
acoustic time is the shortest, the dynamical regime is isobaric and 
the equations of motions are reduced to a 
diffusion-reaction equation in Lagrangian variables\cite{meer93}, and contrary, 
when 
acoustic time is the longest (the long-wavelength limit), the dynamics of RCM 
is described by equations of gas dynamics with equation of state determined by 
``thermal equilibrium'', \ie by balance between radiative cooling and 
external heating\cite{meer87}. 

The above approximations suggest that the wavelength of perturbations
of hydrodynamical variables satisfies one of the two strong inequalities: 
$\tau_A\ll \tau_R$ for the short-wavelength and $\tau_A\gg \tau_R$ for the 
long-wavelength limits, respectively. Perturbations with wavelengths close to 
the characteristic length, \ie $\tau_A\simlt\tau_R$ or $\tau_A\simgt\tau_R$, 
show complex behavior and require in general numerical study treated with the 
full set of equations. We address in
this paper the question of what are the wavelength limits in which isobarical
approximation can be used for description of thermal
instability. We consider perturbations which initially are small (\ie 
linear), and show that these limits are rather broad: at $t>3-15~\tau_A$, 
perturbations with wavelengths much 
larger than wavelengths determined by Meerson 
\etal\cite{meer93}, \ie $\tau_A\gg \tau_R$, evolve isobarically. Our conclusion is
therefore that at comparatively late stages long-wavelength 
perturbations can be described by a diffusion-reaction type equation for 
temperature and density, equivalent to the equation for isobaric 
short-wavelength perturbations. Initial stages of long-wavelength perturbations 
can be understood in terms of wave dynamics in a pressure-free medium, and are
reasonably described by nonlinear advective-diffusion equation for velocity.

The paper is organized as follows: in Section II we present the basic equations 
and discuss limiting cases which allow analytical description, 
in Section III we show numerical results  
for long-wavelength perturbations. Section IV contains a summary.  

\section{ GOVERNING EQUATIONS, CHARACTERISTIC TIMES, LIMITS}
\subsection{Equations}

The basic equations governing dynamics of a radiatively cooling gas were 
proposed originally by Field\cite{field65} and include in the 
energy equation source and sink terms describing heating by external sources 
and radiative energy losses. In planar geometry the equations are
written as 
$${\de \rho \over \de t}+{\de \over \de x}(\rho u)=0,\eqno(1)$$

$$\rho\left({\de u\over \de t}+u{\de u\over \de x}\right)+
{\de P\over \de x}= \eta{\de^2 u\over \de x^2},\eqno(2)$$ 

$${1\over \gamma -1}\left({\de P\over \de t}+u{\de P\over \de x}\right)+
{\gamma\over \gamma-1}P{\de u\over \de x}+\rho{\cal L}(\rho,T)-
{\de \over \de x}\left(\chi{\de T\over \de x}\right)
-\eta\left({\de u\over \de x}\right)^2=0,
\eqno(3)$$
$$P-{{\cal R}\over \mu}\rho T=0,\eqno(4)$$

\noindent
where ${\cal L}$ is the net cooling function 
defined as ${\cal L}(\rho,T)=\rho\Lambda(T)-
\Gamma(\rho,T)$, with $\Gamma$, as a rule, weekly dependent on $\rho$ and $T$, 
other notations have a common meaning. 

\subsection{Characteristic times}

Three characteristic time scales are relevant for description 
of perturbations in RCM: acoustic time $\tau_A=\lambda/c_0$, conductive 
time $\tau_\chi={\cal R}\rho\lambda^2/\mu\chi$, and radiative cooling time 
$\tau_R={\cal R}T/\Lambda\mu\rho$, here $c_0=\sqrt{\gamma P_0/\rho_0}$ is 
the adiabatic sound speed. It is obvious that 

$$\tau_\chi\sim 
3\lambda^2/\ell c_0, $$ 
where $\ell$ is a free-path of gas particles, and
thus in framework of hydrodynamical description, \ie $\lambda\gg \ell$, 

$$\tau_\chi=\lambda \tau_A/\ell\gg \tau_A .$$  
For cooling rate per unit mass $\rho\Lambda(T)\sim \sigma_ic_0\Delta E
n/\mu$, where $\sigma_i$ is the cross-section of inelastic collisions, 
$\Delta E$, the energy lost by particles in a single inelastic
collision, $n$ is the gas number density, characteristic cooling time is 

$$\tau_R={{\cal R}T\over \epsilon\Delta E}{\ell\over c_0},$$
where $\epsilon=\sigma_i/\sigma_e<1$, the ratio of inelastic-to-elastic
cross-sections. Since energy transfered in a single collision, as a 
rule, is small compared to thermal energy $\Delta E< {\cal R}T$, 
we get from here $\tau_R\gg \ell/c_0$. 

\subsection{Limits}

Therefore, three different intervals of wavelengths are relevant, see
Fig. 1: short wavelengths $\lambda_H<\lambda< \lambda_F$, intermediate 
wavelengths $\lambda_F<\lambda<\lambda_{A}$, and long wavelengths 
$\lambda>\lambda_{A}$, here $\lambda_F=\sqrt{\chi_0T_0/\rho_0\Lambda_0}$, 
the Field length determined
as a length where conductive heat transfer and radiative losses equate,
\ie $\tau_R=\tau_\chi$, 
$\lambda_{A}=\tau_R c_0$ is the acoustic length where
$\tau_A=\tau_R$ \cite{meer96}, and $\lambda_H\sim \ell$ is the wavelength 
at which hydrodynamical description violates. It is seen from Fig. 1 that 
always the interrelation between characteristic lengths $\lambda_{H}<
\lambda_F<\lambda_A$ is valid. From this point of view, the ``direct 
crossover'' regime\cite{meer96,meer993} corresponding to the case of large heat 
conduction, $\lambda_A\ll \lambda_F$, is out of the hydrodynamical 
approximation. 

The principal difference between the short and intermediate-wavelength 
intervals is in more efficient conductive heat transfer compared to 
radiative losses for short-wavelength perturbations, which leads
to erasing of a monochromatic perturbation (a normal mode with fixed 
wavenumber $k=2\pi/\lambda$, \eg $\delta T\propto e^{ikx}$) 
with $\lambda < \lambda_F$ 
(see\cite{meer96,field65}). A non-monochromatic localized isobaric perturbation 
(a wave packet of a set of normal modes, \eg $\delta T\propto 
\int \delta T_k e^{ikx} dk$) with 
characteristic size $L<\lambda_F$ shows decrease in amplitude at
initial time, which then changes to increase due to thermal instability
of long-wavelength modes presenting in the perturbation\cite{fer97}. Since 
in these intervals $\tau_A$ is the shortest characteristic time, pressure
balance sets in first over the perturbation 
equations reduce to the isobaric form\cite{dor81,meer89}

$$
{\de T\over \de t}+L(T,P)-{\de \over \de m}\left(\tilde\chi{\de T\over \de
m}\right)=0,\eqno(5)
$$
where 

$$L(T,P)={\gamma-1\over \gamma}{\mu\over {\cal R}}\rho{\cal L}$$
is the reduced cooling function, 

$$\tilde\chi={\gamma-1\over \gamma}{\chi \over PT},$$ 
the reduced heat conduction coefficient, $T dm=dx$, $P=$const. 

In the long-wavelength limit, acoustic time $\tau_A$ is much larger 
than characteristic radiative cooling time $\tau_R$, and conductive 
heat transfer is negligible: $\tau_\chi\sim \lambda^2/\chi
\gg \tau_A\gg \tau_R$, see Fig. 1. In this limit the 
equations can be reduced to more simple form describing gaseous medium with 
a given (anomalous) equation of state\cite{meer87}. 
Indeed, when $\tau_A\gg \tau_R$ the energy balance sets in on the shortest 
time $\tau_R$, and thus temperature and pressure relax rapidly to the values 
determined from the equation of steady energy balance 

$${\cal L}(\rho, T)=0.\eqno(6)$$
Note, that Eq. (6) with cooling function $\Lambda(T)$ shown in Fig. 2 and 
$\Gamma=$const maps $(\Lambda,T)$ plane on $(P,\rho)$ plane to give the  
equilibrium equation of state $P=P(\rho)$ of a van der Waals type. 
Equilibria with temperature in the interval where $d\ln \Lambda(T)/
d\ln T< 1$ are thermally unstable, and correspond to anomalous part of the 
$P(\rho)$-curve with $dP/d\rho<0$\cite{meer96,field65,meer87}. 
These arguments allowed Meerson and Sasorov\cite{meer87} to reduce the momentum 
equation 
of RCM in the long-wavelength limit to the one with a given 
(``equilibrium'') equation of state $P=P(\rho)$ 

$$\rho\left({\de u\over \de t}+u{\de u\over \de x}\right)+
{\de P\over \de x}=\eta{\de^2 u\over \de x^2},\eqno(7)$$ 

\noindent
where $P=P(\rho)$ is determined from Eq. (6). For cooling 
rate with $d \ln \Lambda(T)/d \ln T > 1$ below the 
temperature $T=T_L$ and above $T=T_U$, and $d\ln \Lambda(T)/d \ln T 
< 1$ in the intermediate interval, the equation of state has a form 
similar to the van der Waals equation with two 
rising (stable) branches of the curve $P(\rho)$ in the low density 
$\rho<\rho_U$ (high temperature $T>T_U$) and high density 
$\rho>\rho_L$ (low temperature $T<T_L$) ranges, and a falling 
(unstable) branch of the $P(\rho)$-curve in the intermediate range 
$\rho_U<\rho<\rho_L$ ($T_L<T<T_U$). $N$-shaped curves in Figures 4 and 6 show 
the ``equilibrium'' equation of state $P(\rho)$ corresponding to Eq. (6) 
with the cooling function $\Lambda(T)$ depicted in Fig. 2 
(see, also\cite{meer96,meer993}). 
This equation of state is commonly used for description of pattern formation 
in a bistable medium in the long-wavelength limit\cite{meer96}. (For the 
interstellar 
medium $T_L\sim 10^2$ K and $T_U\sim 10^4$ K, see\cite{spitzer,kaplan}). 

We will mostly concentrate in this paper on the dynamics of perturbations 
in an RCM which is in the upper marginally stable point: 
$T\sim T_U$, $d\ln \Lambda /d\ln T\sim 1$, thus even small (linear) 
perturbations break the initial state. In this state an RCM can be prepared, 
\eg by adiabatically slow decrease of the intensity of 
external heating source in an RCM kept at fixed pressure $P_U$, or adiabatically 
slow increase of pressure at fixed intensity of heating sources. 
As at the equilibrium point $T=T_U$ the compressibility is infinite, \ie
$dP/d\rho=0$, small perturbations around this point evolve to form localized 
structures as was demonstrated by Meerson and Sasorov\cite{meer87}. In the 
next order positive perturbations of density decrease pressure, providing 
anomalous (negative) compressibility 

$$P(\rho)=P_U-{a^2\over 2}\delta\rho^2,\eqno(8)$$

\noindent
where $\delta\rho=\rho-\rho_U$, which amplifies the  
nonlinearity connected with the advective term in momentum equation (3). 
It is clear that nonlinear term in Eq. (8) does not saturate the instability, 
and therefore both the advective term and anomalous negative compressibility 
can be stabilized only by a strong nonlinearity in the equation of state 
$P(\rho)$, which gives rise to the second stable branch. This saturated state 
corresponds  therefore to a new state of a medium separated onto two phases. 
The difference between densities of two phases being in pressure equilibrium 
is considerable, as a rule about two orders of 
magnitude, and only numerical study can be applied to these strongly nonlinear 
saturation stages\cite{meer993}. In order to study dynamics of long-wavelength 
localised structures in a fluid with negative compressibility Meerson and 
Sasorov\cite{meer87} have approximated the falling branch of the equation of 
state as 
$P(\rho)=P_U+{\rm const}/\rho$, and described 
the so-called unlimited instability. 

Initial stages, when $|\delta \rho|\ll \sqrt{2P_U/a^2}$, can be 
described by the advective-diffusion equation 

$$
\rho\left({\de u\over \de t}+u{\de u\over \de x}\right)
=\eta{\de^2 u\over \de x^2}.\eqno(9)
$$ 
Numerical simulations show that equation (9) gives qualitatively reasonable 
description of velocity profiles up to $t\simlt (2-15)\tau_A$ depending on 
wavelength. Within this time interval pressure varies from $P_U$ to minimal 
value $P=P_U- \Delta P_m$, and arised pressure gradients force then velocity 
to steepen sharply. It forms, in turn, a short-wavelength region,
where thermal regime separates the ``equilibrium'' equation of state 
$P=P(\rho)$ and isobaric distribution sets in rapidly, and then velocity 
relaxes gradually to profiles adequate to isobaric dynamics. We argue therefore, 
that at larger times {\it long-wavelength} perturbations in RCM evolve 
isobarically, and can be described by equation (5).  


\section{NUMERICAL RESULTS }


To confirm this conclusion we show here two numerical models which 
follow the evolution of perturbations from initial linear state 
to final asymptotic behaviour. In numerical simulations we have used 
a 6-parametric finite difference scheme of Eqs. (1)--(4) which is free of 
nonphysical effects of numerical viscosity and absolutely stable (the 
detailed description of the scheme will be given elsewhere). 

To make presentation more clear we use dimensionless 
variables with length, time and velocity normalized, respectively, to 
the initial size $\lambda$ of a perturbation (which is assumed to be
localized), to the dynamical time $\tau_A=\lambda/c_0$, and to the 
adiabatic sound speed $c_0$; hydrodynamical variables $\rho,~P$ and $T$ are 
normalized to their unperturbed values. In these units cooling time is equal to 
$\tilde\tau_R=\tau_R/\tau_A$. As we concern here
dynamics of long-wavelength perturbations, the characteristic size 
is $\lambda> \lambda_F$, (in numerical examples given in the paper 
$\lambda_F=0.001\lambda$). Dimensionless cooling function $\Lambda(T)$ 
has been taken in the form 

$$
\Lambda(T)=\Bigl\{\alpha_1+\alpha_2 \tan 
[{\pi\over 2}(T^\ast-T)]\Bigr\}^{-1},\eqno(10) 
$$
as shown in Fig. 2. This function represents typical cooling function
for bistable media, and generates equation of state of the van der Waals 
type when ``thermal equilibrium'', \ie $\rho\Lambda(T)=\Gamma(\rho,T)$ is
reached; we assume through the paper $\Gamma(\rho,T)=\Lambda(1)$=const.
The unperturbed state was taken near the upper unstable point. We present
here results with initial conditions taken in the form of small 
velocity perturbations corresponding to a localized symmetric inflow to
the center of a grid zone 

$$
u(x,0)=\cases{-u_0\sin[k(x-x_0)],~|x-x_0|\leq 2\pi/ k,\cr 
~~0,~\qquad ~\qquad~\qquad\rm otherwise,}\eqno(11)
$$
while perturbations of $\rho$, $T$, and $P$ were zero. This choice of initial 
conditions seems to be most appropriate in study of long-wavelength 
perturbations. Due to radiative losses initial conditions corresponding to 
perturbations of density, temperature and pressure are inevitably 
more affected by the procedure of preparation such perturbations. 
We therefore, do not consider here alternative dynamics which would start from 
small perturbations of density, temperature and pressure.

For adopted type of initial conditions velocity perturbations disturb 
density, temperature and pressure,  
and radiative cooling drives them to relax in a short time 
to values determined by thermal equilibrium (6). Arised gradients of
pressure redistribute then velocities, such that the resultant
velocity field tends to set the flow in isobaric regime. At 
late stages the flow is approximately isobaric through the perturbed
region. The boundary conditions used correspond to a ``closed'' box, \ie
zero velocity and gradients of density and pressure. In most 
models the grid zone size was 20 times of the size of a localized 
perturbation, however test runs with larger 
size (80 times of the perturbation) show results to be insensitive 
to the grid size. This is a direct consequence of the fact that the 
compressibility in the unperturbed state (which is taken to be close to the 
upper unstable point) is very high, and a localized 
perturbation propagates slowly: the acoustic velocity 
$c_U=\sqrt{(dP/d\rho)_U} \simeq 0.25$. Moreover, the initial perturbation
brings most fraction of perturbed gas into a state with negative
compressibility where the perturbation does not propagate. 
The time required for perturbations to reach a boundary is therefore 
at least 80 dynamical times, which is larger than the total 
computational time in all cases we considered. This means, that the 
boundary conditions we used weakly affect dynamics of perturbations. 
(Note, that systems whose unperturbed states lie on the low-density 
stable branch well below the upper unstable point, can be brought into thermal 
instability only by perturbations of finite, \ie not small, amplitude. We do 
not consider such cases in this paper.)

Fig. 3 shows dynamics of a perturbation with initial amplitude of
velocity $u_0=0.03$ and with $\tilde\tau_R=0.33$ (\ie $\lambda=3\lambda_A$),
thus it corresponds to the long-wavelength limit. On the upper panel velocity 
profiles  are plotted. At initial times, $t\leq 10$, as gas in central regions 
is involved in motion the velocity amplitude decreases. In gases with 
normal equation of state, \ie $dP/d\rho>0$,
perturbation decays onto two acoustic waves propagating in opposite directions, 
and a standing and damping entropy mode 
[in case of perturbations given by eq. (11), decayed waves represent 
two pieces of sinusoidal waves with velocity amplitude of $u_0/2$]. 
Radiative losses with $\tilde\tau_R\ll 1$ modify the inflow, transfering
the internal (compressed) gas onto the anomalous part of equation of
state ($dP/d\rho<0$), and thus make the perturbation locally growing,
\ie non-decaying. At times $10<t<15$ the velocity profile steepens
and motion concentrates in a rather narrow region with increasing
density, as shown in the bottom panel. 

Relaxation of pressure to the equilibrium (anomalous) equation of state 
$P(\rho)$ occurs at times $t\sim \tilde\tau_R$ (these stages are not 
shown). Increasing density in central regions amplifies pressure gradient 
through the perturbation, and thus forces the velocity amplitude to increase. 
It results in more sharp steepening of velocity profiles compared to the
pressure-free case as seen in Fig. 3a for time interval $10<t<15$: 
the profile for $t=15$ clearly shows two inflection points outside the 
central region very close to peak values, which is not what can be
expected for pressure-free motions, or motions with normal equation of
state. In turn, the inflow with increasing velocity amplitude forms very 
narrow region (a ``droplet'' with sizes $\Delta x\ll \lambda$, as seen in 
Fig. 3b), through which pressure equilibrium sets in rapidly. 
As a result, at these stages structure of the perturbation can be
understood as distinguished onto two regions: first, the
``long-wavelength'' part outside the ``droplet'' where the equilibrium  
equation of state dominates, and second, the ``short-wavelength'' part 
(the ``droplet'' itself) where motion tends to set in the pressure 
equilibrium. The two regimes manifest themselves clearly in Fig. 4, 
where loci $P-\rho$ connecting pressure and
density through the computational zone for different times, 
\ie $P(x,t)$ and $\rho(x,t)$ profiles with eliminated $x$, 
are plotted: the short piece of line at the upper equilibrium point shows 
$P-\rho$ locus for $t=10$. The locus corresponding to $t=15$ is readily
seen to consist of two parts: a small piece of a $P-\rho$ curve starting from 
the unperturbed state with decreasing pressure and density almost
parallel to the {\it growing} branch of the equation of state 
(corresponding to external, long-wavelength regions of the perturbation 
where density and pressure decrease), which then  
turns around and goes with increasing density and almost constant pressure 
(this part corresponds to internal, short-wavelength regions).
Thus, at these stages pressure {\it separates} the equilibrium curve
$P(\rho)$, and subsequent evolution is characterized by settling the
overall flow down to isobaric regime. 
At $t\simgt 20$ most fraction of density perturbation concentrates in the
isobaric ``droplet'' with characteristic size of $|x-x_0|=0.2$ and in 
the conductive interface, which are in pressure
equilibrium with gas outside (variation in pressure through the
perturbed region is less than 0.001, and main contribution comes from 
surrounding regions where density and pressure lie on the equilibrium
equation of state). (Strongly speaking, in models with 
``closed'' box boundary conditions we used, pressure depends on time. 
However, as mentioned above, in all considered cases asymptotic behaviour of 
perturbations establishes on time interval shorter than the time required 
for a perturbation to reach boundaries. One can think, therefore, 
that isobaric regimes asymptotically established in numerical models 
with ``closed'' box boundaries give qualitatively correct description 
of the situation in systems with free boundary and fixed external pressure.) 

After the short-wavelength droplet is formed, velocity decreases to $u< 0.01$
with profiles corresponding to a weak inflow, so more gas continues to
increase mass of the droplet. Subsequent evolution
follows the isobaric regime with $P-\rho$ loci practically coincident 
with that reached at $t=20$, as shown in Fig. 4. Note that in models with 
fixed equilibrium equation of state $P=P(\rho)$, 
see\cite{meer96,meer993}, the established inflow velocity is, in general, 
higher than in our calculations with energy equation (3) solved explicitly. 
This is due to the fact that in models with the 
equilibrium equation of state, $P=P(\rho)$, pressure reaches minimal value 
in the interface layer, and thus at equal conditions (\eg viscosity and 
conductivity) arised pressure gradients generate higher acceleration. 

In Fig. 5 and 6 we show results for a perturbation with larger spatial size: 
$\lambda=30\lambda_A$, \ie  $\tilde\tau_R=0.033$. Initial conditions are
same as in the previous case. The perturbation is seen to reveal similar 
behavior as that of smaller size, however, the dynamics is more
violent. The difference is mainly in time
scales and the amplitude of velocity: perturbations of longer wavelength 
develop higher velocities and pass same stages quicker. The enhanced
steepening of the velocity profile due to negative compressibility
occurs at $t\simlt 2$. Here the difference between enhanced and
normal steepening is more obvious than in previous case: rather flat
increase in velocity in outer regions, and very sharp growth near the
symmetry plane. At $t\sim 3$ the inflow with amplitude
$u\simeq 0.08$ forms a narrow short-wavelength droplet with almost
constant pressure inside [as seen from $P-\rho$ loci on $(P,\rho)$
plane, Fig. 6]. However, as the velocity amplitude developed is large, 
pressure in the droplet is higher than the unperturbed value by $\Delta P
\sim \rho u^2\sim 0.01$, and in very central parts of the droplet 
overpressure reaches the value $\Delta P\sim 0.03$ due to higher
densities [note, that $P-\rho$ locus in the central overpressured region 
goes along the equilibrium equation of state $P(\rho)$ corresponding to the 
second stable brach]. On next stages, $3<t\simlt 4$, the
overpressure leads to expansion of the short-wavelength region (with
almost isobaric pressure distribution inside) to relax to the equilibrium
with external regions. This relaxation occurs more violent than in
the previous case, due to larger variations in pressure, $|\Delta P| 
\sim 0.01-0.03$ around the unperturbed state. As a result, subsequent 
growth of gas mass in the short-wavelength (droplet) region proceeds in
a damped oscillatory regime, as seen on $(P,\rho)$ plane in Fig. 6. 

The dependence of dynamics of perturbations
on wavelength (or equivalently on $\tau_R$) can be understood 
qualitatively from simple estimations of steepening of velocity
profiles. The characteristic formation time of a short-wavelength region
(a droplet) due to steepening can be estimated from $at_d^2\sim 2\lambda$, 
where $a$ is the acceleration caused by the negative compressibility, 

$$a\sim {|\nabla P_m|\over \rho_0}\sim { \Delta P_m\over 
\rho_0\lambda},\eqno(12)$$ 
$\Delta P_m$ is the maximal 
deviation of pressure from the unperturbed value due to density
increase. This gives in dimensionless units 

$$ 
t_d\sim \sqrt{{2\over \Delta P_m}}.\eqno(13) 
$$
The deviation of pressure from the unperturbed value increases along the
equilibrium $P(\rho)$-curve until radiative cooling time $\tau_R$ is
less than the acoustic time for the shortest wavelength 
$\lambda_s$ generated due to steepening. Thus, $\Delta P_m$ can be determined 
from the condition $\tilde\tau_R=\lambda_s$. 
To estimate $\lambda_s$ we assume that perturbation of density caused by
initial inflow (11) in dimensionless units is $\delta\rho_0(x)\sim 
u_0(x)$. Then, due to thermal instability it grows as $\delta\rho(x,t)
\sim u_0\exp(t/\tau_R)$. On the other hand, one can expect that 
$\lambda_s\sim \lambda\exp(-t/\tau_R)$, thus giving $\delta\rho\sim u_0
/\lambda_s$. Substituting here $\Delta P_m\sim c_A^2\delta \rho$, where
$c_A^2=|dP/d\rho|$ estimated on the anomalous branch of equation of
state [$c_A^2\sim 0.1$ at $\rho\sim 1.5$] 

$$
\Delta P_m= {u_0c_A^2\over \lambda_s},\eqno(14)
$$
We obtain therefore for $t_d$:

$$
t_d\sim \sqrt{{2\tilde\tau_R\over u_0c_A^2}},\eqno(15)
$$
which provides with reasonable accuracy an estimate for the time 
when the numerical solutions turn from the regime with the equilibrium equation 
of state, $P(\rho)$, to the isobaric regime. 

It is clear from these estimates
and from numerical results presented, that at small values of the
ratio $u_0c_A^2/\tilde\tau_R$ (less than 0.1) long-wavelength pertubations 
develop quiescently as isobaric. In dimensional units the condition for
long-wavelength perturbations to evolve in a quiescent isobaric regime
is written as 

$$
{\tau_A\over \tau_R}\simlt 0.1 {c_0\over u_0}\left({c_0\over
c_A}\right)^2.\eqno(16)
$$

When $u_0c_A^2/\tilde\tau_R>0.1$, high pressure gradients developed at initial 
``equilibrium'' stages [\ie along $P(\rho)$ equation of state] give rise to 
very fast steepening of velocity profiles, and as a result to formation of 
the overdense and overpressured central short-wavelength 
droplet, which then relaxes to the equilibrium in an oscillatory regime with
variations in pressure of $|\Delta P| \simlt 0.03$ around the
unperturbed value. Regimes with the equilibrium 
equation of state, $P(\rho)$, can exist for long-wavelength
perturbations only at initial stages, $t\simlt t_d$. Subsequent 
evolution implies generation of a short-wavelength core, which in turn
results in separation of pressure of the equilibrium curve $P(\rho)$. 
Formation of such a short-wavelength region is an inevitable consequence
of the ``unlimited'' instability when narrow localized structure
develops in a medium on the anomalous branch of the equilibrium equation
of state$^4$. In Fig. 7 we show the dependence of pressure in
the central droplet on time for several values of
$\tilde\tau_R$ and fixed $c_A^2=0.1$ at $\rho=1.5$. The amplitude of 
pressure oscillations is clearly seen to increase for smaller
$\tilde\tau_R$: approximately, at initial times, when the central
droplet follows the equilibrium equation of state,  
$\Delta P_0\sim \Delta P_m\sim u_oc_A^2/\tilde\tau_R$. The period of
oscillations is ${\cal T}\sim \lambda_s\sqrt{\rho_d/\Delta P}$, where $\rho_d=5$
is the density on the second stable branch; at initial times 
${\cal T}\sim 2.24 \tilde\tau_R^{3/2}/\sqrt{u_0c_A^2}$, however then it
grows due to damping of $\Delta P$. 

\section{SUMMARY}

We have considered nonlinear evolution of initially small 
(linear) perturbations in 
a radiatively cooling medium in the long-wavelength limit, \ie when 
acoustic time is larger than radiative cooling time. We have not been 
assuming pressure and density to satisfy the equilibrium equation of
state, but studied the problem numerically with the radiative energy loss 
term in energy equation treated explicitly. 
For initial conditions in the form of small localized velocity
perturbation around a uniform equilibrium close to the upper unstable
point, we have shown that early stages of evolution 
tend to proceed along the equlibrium equation of state, 
$P(\rho)$, with velocity profiles to steepen as in pressure-free case. 
At intermediate stages density increases enough and anomalous (negative) 
compressibility becomes important and results in 
more sharp steepening of velocity profiles, 
which in turn forms a narrow (short-wavelength) region where pressure 
separates the value determined by the equilibrium equation of state and 
sets in approximate isobaric distribution rapidly. One may say that 
thermal instability works as a regulator which prevents large pressure 
gradient to arise. The maximal amplitude of pressure developed 
in the perturbation when it separates the equilibrium curve $P(\rho)$, 
is proportional to the amplitude of initial velocity perturbation and 
inversely proportional to the radiative cooling time [as described by 
Eq. (14)] 

$$\Delta P_m\sim P_0{u_0\over c_0}{\tau_A\over \tau_R}
\left({c_A\over c_0}\right)^2.\eqno(17)$$
At late stages, gas in the short-wavelength region 
reaches the high-density (stable) equilibrium branch of the equation of 
state, and subsequent evolution corresponds to the isobaric inflow of 
low-density gas onto the high-density core (the droplet). 
We have shown thus, that nonlinear evolution of long-wavelength
perturbations after the time interval 

$$
t_d\sim {c_0\over c_A}\sqrt{2\tau_A\tau_R
{c_0\over u_0}},\eqno(18)
$$
reaches quiescent isobaric regime, if the wavelength is restricted from
above as 

$$
\lambda< 0.1\lambda_A{c_0\over u_0}\left({c_0\over c_A}\right)^2.\eqno(19)
$$
In these limits, long-wavelength perturbations in thermally unstable 
media can be described as isobaric by the reduced equation (5). Perturbations 
with wavelengths longer than the limit (19), after $t=t_d$ when pressure
separates the equilibrium equation of state, develop localized oscillating 
structures with initial pressure amplitude determined by Eq. (14) and 
damped then due to viscosity. 

\acknowledgments

YS acknowledges financial support from Russian Foundation of Basics 
Research (grant 94-02-05016-a), and from Italian Consiglio Nazionale 
delle Ricerche within NATO Guest Fellowship programme 1996 (Ann. No. 
219.29). We acknowledge anonimous referees for critical remarks. 
\bigskip
\parindent=0pc
\vfill\eject

\vskip 1truecm
\noindent
\smallskip

\vfill\eject


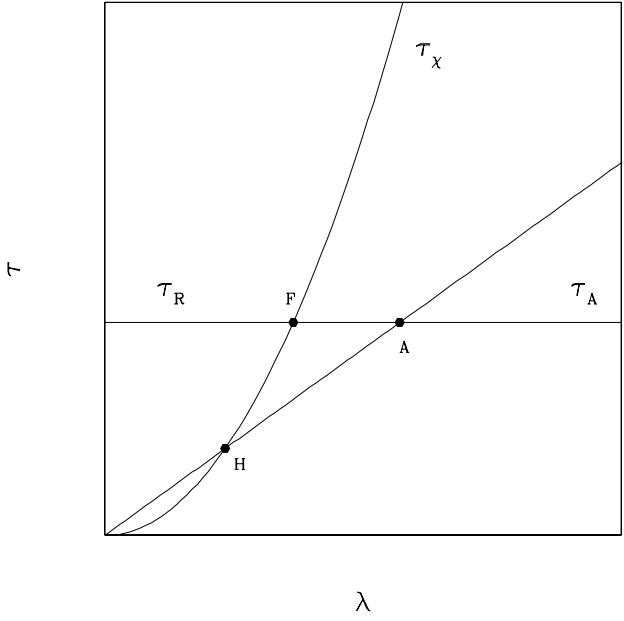
\begin{figure}
\caption{
The dependence of characteristic times on wavelength. 
Point $F$ corresponds to the Field length $\lambda_F$, $A$ --- to the 
acoustic thermal length $\lambda_A$, $H$ --- to the length where 
hydrodinamical description violates $\lambda_H\sim \ell$. The interval 
$\lambda_H<\lambda<\lambda_F$ corresponds to the short-wavelength limit,
$\lambda_F<\lambda<\lambda_A$ --- to the intermediate wavelengths, and 
$\lambda_A<\lambda$ --- to the long-wavelengths. 
}
\end{figure}

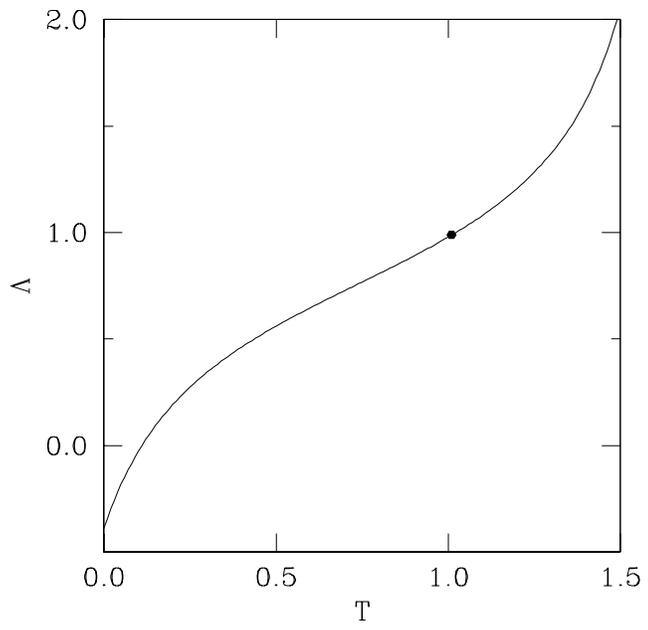
\begin{figure}
\caption{
The cooling function $\Lambda(T)$ used in the paper:
$\alpha_1=0.915$, $\alpha_2=0.61$, $T^\ast=1.11$. 
The point indicates the unperturbed state; the heating rate was taken to 
balance energy losses in this point. 
}
\end{figure}

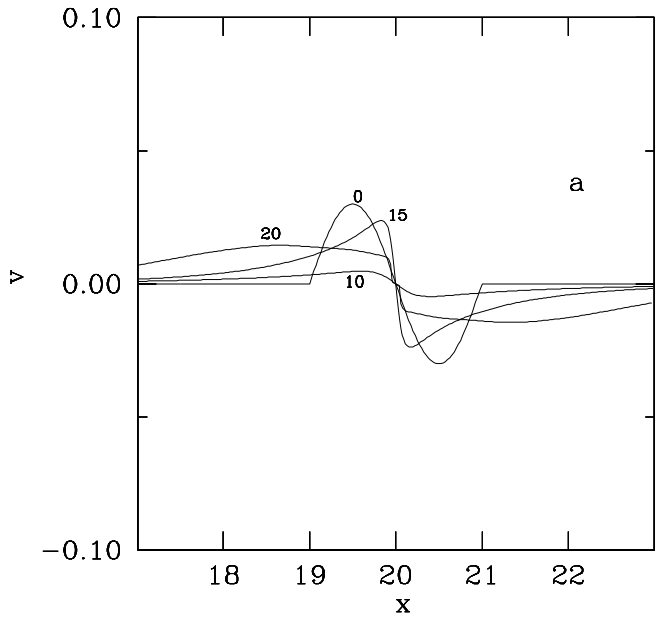
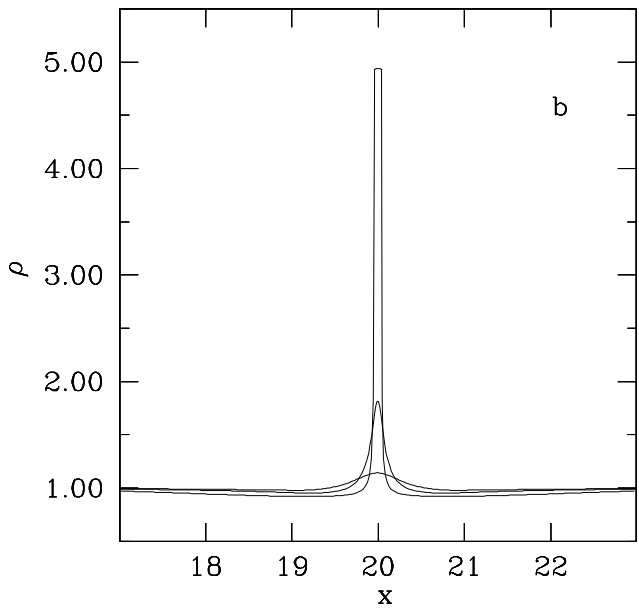
\begin{figure}
\caption{
Nonlinear evolution of perturbations in RCM: $\lambda=3
\lambda_A$,
$\tilde\tau_R=0.33$. Upper
panel (a) shows velocity profiles for the times 
$t=0,~10,~15,~20$. The
amplitude of initial velocity perturbation is $u_0=0.03$. Density,
temperature and pressure have not been perturbed.  Bottom panel (b) shows
density profiles corresponding to the velocity profiles given on the
upper panel [initial density profile $\rho_0(x)$=const is not
shown]. 
}
\end{figure}

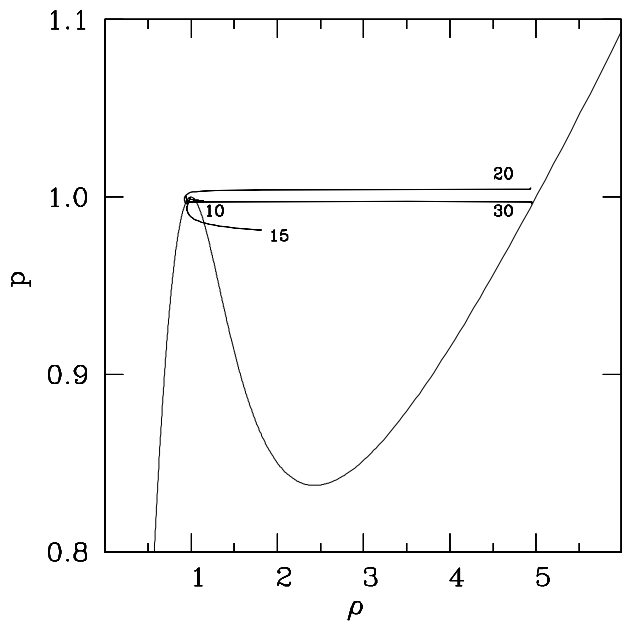
\begin{figure}
\caption{
$P-\rho$ loci [obtained from $P(x,t)$
and $\rho(x,t)$ profiles with eliminated $x$] on the $(P,\rho)$ plane,
where the graph of the equilibrium equation of state $P(\rho)$ is also 
ploted. The unperturbed state corresponds to a point at the $P(\rho)$ 
curve left to the upper stable point. Curves for $t=5,~10,~15,~20,~30$ 
are shown. The short piece (10) of line near this
point represents (splitting) $P-\rho$ loci for times $t=5$ and 10. 
}
\end{figure}

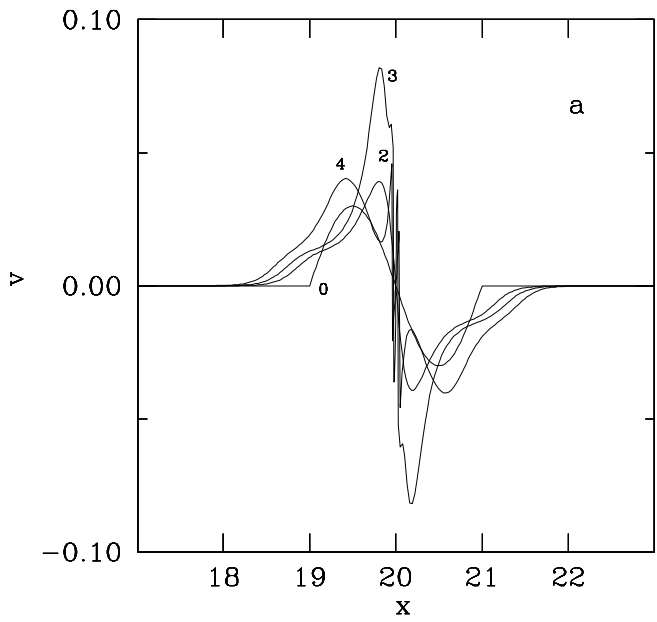
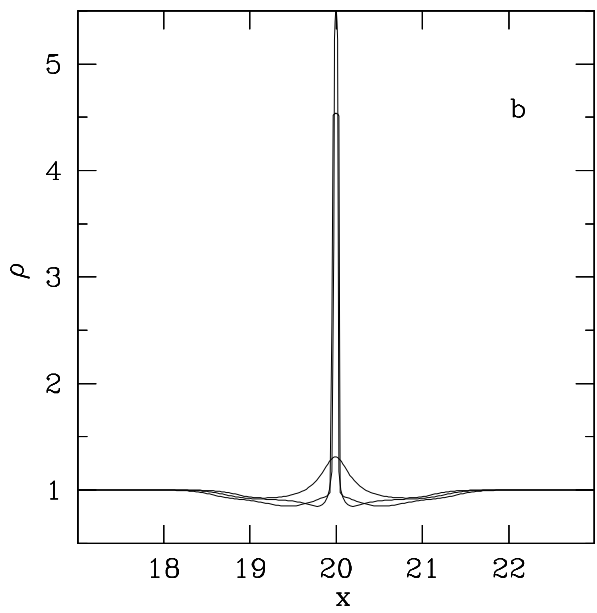
\begin{figure}
\caption{
Same as in Fig. 3 for $\lambda=30\lambda_A$,
$\tilde\tau_R=0.0033$, $u_0=0.03$. Panel (a) shows velocity profiles
for times $t=0,~2,~3,~4$. 
Panel (b) shows density profiles for
times $t=2, ~3,~4$. 
}
\end{figure}

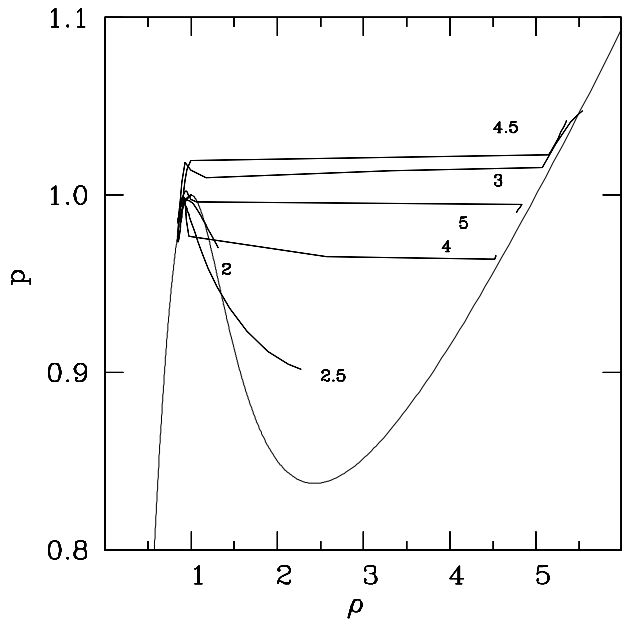
\begin{figure}
\caption{\hbox to 18cm{ 
$P-\rho$ loci for the case shown in Fig. 5: curves are
shown for $t=2,~2.5,~3,~4,~4.5,~5$.
\hfill}}
\end{figure}

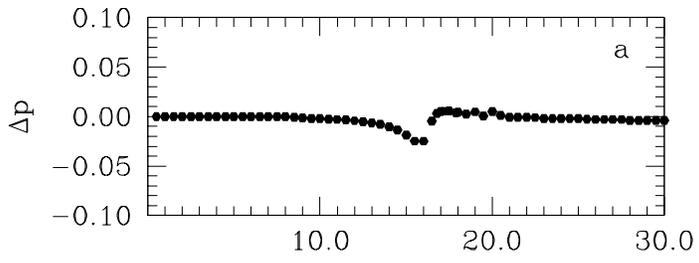
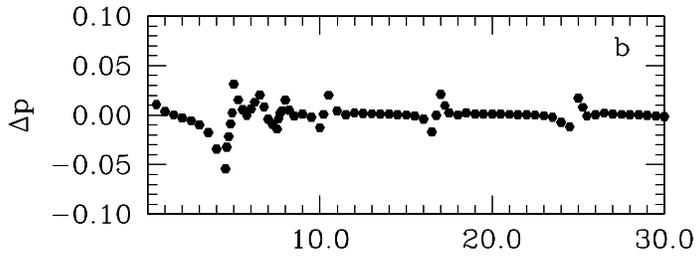
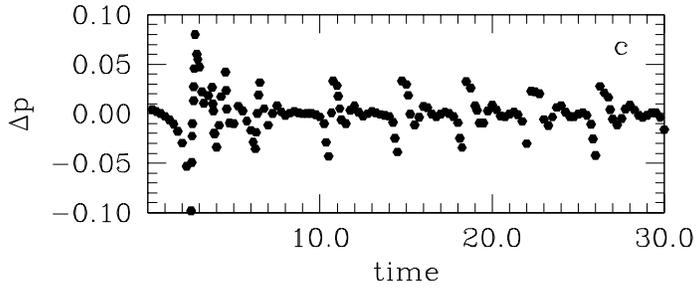
\begin{figure}
\caption{
$P_c(t)$ for several values of $\tilde\tau_R$; plots show
$\Delta P(t)=P_c(t)-1$: $\tilde\tau_R=0.33$ (a), $\tilde\tau_R=0.1$ (b)
and $\tilde\tau_R=0.033$ (c). At $t<t_d$ $P_c(t)$ monotonously decreases
along $P(\rho)$ curve to minimal value, at $t\sim t_d$ it separates 
$P(\rho)$ value and grows sharply to relax then to the isobaric
distribution. Time is given in units $\tau_A$. 
}
\end{figure}


%

\begin{references}
\bibitem{meer96}B. Meerson, Rev. Mod.\  Phys.\ {\bf
68}, 215 (1996).
\bibitem{field65}G. B. Field, Astrophys. J. {\bf 142}, 531 (1965).
\bibitem{dor81}A. G. Doroshkevich, Ya. B. Zeldovich, 
     Zh. Eksp. Teor. Fiz. {\bf 80}, 801 (1981) 
     [Soviet. Phys. JETP {\bf 53}, 405 (1981)]. 
\bibitem{meer87}B. I. Meerson, P. V. Sasorov, 
     Zh. Eksp.Teor. Fiz. 
     {\bf 92}, 531 (1987) 
     [Sov.Phys. JETP {\bf 65}, 300 (1987)]. 
\bibitem{meer93}B. Meerson, E.R. Priest, and C.D.C. Steel, Geophys. Astrophys. 
     Fluid Dyn. {\bf 71}, 243 (1993). 
\bibitem{meer993}B. Meerson, C.D.C. Steel, A.M. Milne, and E.R. Priest, 
     Phys. Fluids B {\bf 5}, 3417 (1993). 
\bibitem{fer97}A. Ferrara, and Yu. Shchekinov, 
     Geophys. Astrophys. Fluid 
     Dynamics, {\bf 84}, 273 (1997)
\bibitem{meer89}B. Meerson, 
     Astrophys. J., {\bf 347}, 1012 (1989)
\bibitem{spitzer}L. Spitzer, Jr., {\it Physical Processes in the 
     Interstellar Medium} (Wiley-Interscince, New York, 1978).  
\bibitem{kaplan}S.A. Kaplan, and S.B. Pikelner, {\it Physics of the 
     Interstellar Medium} (Nauka, Moscow, 1979).  
\end{references}
\end{document}